\documentclass[amssymb,aps,prb,twocolumn,showpacs]{revtex4-1}

\usepackage{latexsym,amsmath,graphics,graphicx,cmbright,exscale,colordvi,accents,lipsum}
\usepackage[T1]{fontenc}
\usepackage{color}
\usepackage[normalem]{ulem}
\usepackage{tikz}
\usepackage[colorlinks=true,linkcolor=blue,citecolor=blue,urlcolor=blue]{hyperref}  

\usepackage{dcolumn}
\usepackage{booktabs}       
\usepackage{multirow}

\newcommand{\ie}{\textit{i.e.}}
\newcommand{\eg}{\textit{e.g.}}
\newcommand{\cf}{\textit{cf.}}
\newcommand{\etal}{\textit{et al.}}

\newcommand{\beq}{\begin{equation}}
\newcommand{\eeq}{\end{equation}}
\newcommand{\bea}{\begin{eqnarray}}
\newcommand{\eea}{\end{eqnarray}}
\newcommand{\vc}[1]{\vec{#1}}                       
\newcommand{\mrm}[1]{\mathrm{#1}}                   

\newcommand{\EF}{E_\mrm{F}}                         
\newcommand{\sunit}[2]{#1~#2}                       
\newcommand{\Angs}{\mbox{\r{A}}}                    
\newcommand{\eV}{\mrm{eV}}                          
\newcommand{\meV}{\mrm{meV}}                        
\newcommand{\mB}{\mu_\mrm{B}}                       

\renewcommand{\Im}{\mrm{Im}\,}
\newcommand{\imagi}{\mathsf{i}\,}                   
\newcommand{\upa}{\uparrow}
\newcommand{\dna}{\downarrow}

\begin{document}

\title{Spin excitations in 3d transition-metal adatoms on Pt(111):\\
       Observable with inelastic scanning tunneling spectroscopy or not?}
\author{Benedikt Schweflinghaus}
\author{Manuel dos Santos Dias}
\author{Samir Lounis}\email{s.lounis@fz-juelich.de}
\affiliation{Peter Gr\"unberg Institut and Institute for Advanced Simulation, Forschungszentrum J\"ulich, 52425 J\"ulich \& JARA, Germany}

\date{\today}

\begin{abstract}
  Spin excitations in atomic-scale nanostructures have been investigated with inelastic scanning tunneling spectroscopy, sometimes with conflicting results. 
  In this work we present a theoretical viewpoint on a recent experimental controversy regarding the spin excitations of Co adatoms on Pt(111). 
  While one group [Balashov \etal, Phys. Rev. Lett. \textbf{102}, 257203 (2009)] claims to have detected them, another group reported their observation only after the hydrogenation of the Co adatom [Dubout \etal, Phys. Rev. Lett. \textbf{114}, 106807 (2015)]. 
  Utilizing time-dependent density functional theory in combination with many-body perturbation theory we demonstrate that, although inelastic spin excitations are possible for Cr, Mn, Fe, and Co adatoms, their efficiency differs. 
  While the excitation signature is less pronounced for Mn and Co adatoms, it is larger for Cr and Fe adatoms. 
  We find that the tunneling matrix elements related to the nature and symmetry of the relevant electronic states are more favorable for triggering the spin excitations in Fe than in Co. 
  An enhancement of the tunneling and of the inelastic spectra is possible by attaching hydrogen to the adatom at the appropriate position.
\end{abstract}

\maketitle

\section{Introduction}
\label{sec:Intro}

Inelastic scanning tunneling spectroscopy\cite{LothNewJPhys10,LorenteGauyacqPRL09,LounisSurfaceScience14} (ISTS) is the tool of choice to probe spin excitations (SEs) in nanostructures down to the single atom limit.\cite{HirjibehedinScience07,BalashovPRL09,OtteNP08,KhajetooriansPRL11,ChilianPRB11,KhajetooriansPRL13,Rau2014,DuboutPRL15}
Characterization and understanding of the properties of SEs in nanosized magnets are prerequisites for the development of new magnetic storage and logic elements, which rely on fast dynamical magnetization processes.\cite{KhajetooriansScience11,Loth2012,KhajetooriansNP12}

Concomitant with the experimental advances, controversies arose on the interpretation of ISTS measurements, \eg, those of Co adatoms deposited on the Pt(111) surface. 
For instance, Balashov \etal\cite{BalashovPRL09} reported SE signatures for Co and Fe adatoms on Pt(111). 
As an external magnetic field was not available, the magnetic character of the excitations was proposed by comparison with the optical measurements of the magnetic anisotropy energy performed by Gambardella \etal\cite{GambardellaScience03} 
Khajetoorians \etal\cite{KhajetooriansPRL13} revisited this problem. 
They found SEs in Fe adatoms at much lower energies than reported in Ref.~\onlinecite{BalashovPRL09} and could not detect SEs for Co adatoms.\cite{privateComm} 
The experiments of Dubout \etal\cite{DuboutPRL15} support these latter findings, and in addition demonstrate that contamination with hydrogen leads to enhanced SE signatures. 
However, after exposure to hydrogen, the chemical composition of the $\mrm{CoH}_x$ complexes is uncertain.

We address this controversy by investigating the electronic structure of Cr, Mn, Fe and Co adatoms deposited on Pt(111), how the SEs arise from it, and finally how they impact the ISTS signal. 
We make use of our recently developed methodology\cite{SchweflinghausPRB14} that combines time-dependent density functional theory\cite{TDDFT_RungeGrossPRL84} (TDDFT) with many-body perturbation theory\cite{FetterWaleckaBook71,ZhukovPRB06} (MBPT). 
We demonstrate that, although SEs are possible in all these magnetic adatoms, their efficiency is dramatically affected by the tunneling matrix elements. 
Contingent on the details of the electronic structure and its symmetry, the decay into vacuum changes considerably from adatom to adatom. 
We find that Fe is the most favorable candidate for probing SEs with ISTS, and show how hydrogen can deeply alter the electronic structure and enhance the inelastic contribution to the tunneling current.

Our paper is organized as follows: 
in Sec.~\ref{sec:Method} we outline our theoretical approach. 
We then present in Sec.~\ref{sec:PureAdatomSpectra} results for the SEs of 3$d$ adatoms and the corresponding inelastic spectra, followed by the role of hydrogen in Sec.~\ref{sec:HydrogenatedAdatomSpectra}, focusing on $\mrm{FeH}$ and $\mrm{CoH}$ dimers. 
Our conclusions are collected in Sec.~\ref{sec:Summary}.

\section{Theoretical framework}
\label{sec:Method}

Our investigations are carried out by combining two methods based on density functional theory: 
($i$) structural optimization is performed with the projector augmented wave (PAW) method\cite{Blochl1994} as implemented in the Vienna \textit{Ab initio} Simulation Package\cite{VASP_KresseHafnerPRB93,VASP_KresseFurthmuellerPRB96} (VASP) and 
($ii$) the ground state electronic and magnetic properties as well as the SEs spectra are evaluated with the Korringa-Kohn-Rostoker (KKR) Green function method.\cite{KKR_PapanikolaouJPCM02} 
The tunneling spectra were extracted considering the Tersoff-Hamann approximation,\cite{TersoffHamannPRL83} which demands the evaluation of the local density of states (LDOS) in the vacuum at the position of the ISTS probe tip. 
In the following two subsections we will describe how the electronic structure is computed and how we access the SEs in the adatoms and their signature in the vacuum region.

\subsection{Electronic structure: a multi-code approach}

The structural optimization calculations are carried out with the VASP program,\cite{VASP_KresseHafnerPRB93,Blochl1994,VASP_KresseFurthmuellerPRB96} utilizing the generalized gradient approximation for the exchange-correlation potential.\cite{PBE_PerdewPRL96} 
For the considered $\mrm{FeH}$ dimers an inversion symmetric 3$\times$3 supercell is set up, consisting of five Pt layers as well as two $\mrm{FeH}$ dimers, that are placed in the layers above and below the slab. 
The lattice constant is set to $a=\sunit{3.924}{\Angs}$. 
The height of the unit cell is $\sunit{28}{\Angs}$, such that the distance from the impurity atom to the slab of the neighboring unit cell is about $\sunit{15}{\Angs}$. 
The Brillouin zone is sampled with a 7$\times$7$\times$1 k-point mesh. 
An energy cutoff parameter of $\sunit{500}{\eV}$ is chosen and the tolerance for the forces is set to $\sunit{0.01}{\eV/\Angs}$.

For the KKR-based simulations,\cite{KKR_PapanikolaouJPCM02} we consider a Pt(111) slab of 22 layers, capped by vacuum regions equivalent to four Pt layers, above and below the slab. 
Also here, the lattice constant is set to $a=\sunit{3.924}{\Angs}$. 
We adopt the atomic sphere approximation (ASA) considering the full charge density with an $\ell_\mrm{max}=3$ cutoff and the local spin-density approximation (LSDA).\cite{LSDA_VoskoWilkNusairCJP80} 
Each single adatom, with or without an additional hydrogen atom next to it, is embedded in real-space on the top surface of the slab, considering a inward relaxation of 20\% of the Pt bulk interlayer distance. 
The two-dimensional (2D) Brillouin zone is sampled with a 180$\times$180 k-point mesh. 
The energy integrations are performed with 60 points on a rectangular contour in the complex plane,\cite{Wildberger1995} including 5 Matsubara frequencies with $T=\sunit{50.26}{K}$. 
Throughout this paper the focus is on 3$d$ adatoms on the fcc stacking site, since the hcp stacking site leads to similar findings regarding the role of hydrogen.\cite{KhajetooriansPRL13}

\subsection{Spin excitations: renormalized electronic structure}

In order to describe SEs we apply a method that has been established in recent years.\cite{LounisPRL10,LounisPRB11,SchweflinghausPRB14,dosSantosDiasPRB15} 
In a first step it allows the calculation of the intrinsic SE spectra, described by the transverse dynamical susceptibility $\chi^{+-}$, measuring the probability of lowering the spin of the nanostructure by $\hbar$. 
It is computed via TDDFT\cite{LounisPRL10,LounisPRB11} after solving the Dyson-like equation
\bea
      \chi^{\sigma\overline\sigma}(\omega)
  =   \chi_0^{\sigma\overline\sigma}(\omega)
    + \chi_0^{\sigma\overline\sigma}(\omega)\,U\,\chi^{\sigma\overline\sigma}(\omega) \;,
  \label{eq:DysonLikeEq}
\eea
where
\bea
      \chi_0^{\sigma\overline\sigma}(\omega)
  &=& -\frac{1}{\pi}\!\int^{\EF}\!\!\!\!\text{d}{E}\,
      \Big(G_0^{\bar\sigma}(E+\omega+\imagi 0^+)\,\Im\,G_0^{\sigma}(E) \nonumber\\
  & & + \Im\,G_0^{\bar\sigma}(E)\,G_0^{\sigma}(E-\omega-\imagi 0^+)\Big)
  \label{eq:Chi}
\eea
is the Kohn-Sham susceptibility, which describes the creation of electron-hole excitations (or Stoner excitations) of opposite spin, leading to the damping of the SEs. 
It is connected to $\chi$ via the frequency-independent exchange-correlation kernel, $U$, in the adiabatic LSDA\cite{GrossKohnPRL85,LiuVoskoCanJP89} (ALSDA). 
The Green functions $G_0^\sigma$ are provided directly by the KKR method,\cite{KKR_PapanikolaouJPCM02} and are utilized after being projected on a local basis as described elsewhere.\cite{LounisPRL10,LounisPRB11} 
Here, $\chi^{\upa\dna}$ and $\chi^{\dna\upa}$ correspond to $\chi^{+-}$ and $\chi^{-+}$, respectively.\cite{SchweflinghausPRB14} 

Once the susceptibility is calculated, one can explore the intrinsic properties of the SEs.\cite{dosSantosDiasPRB15} 
However, the evaluation of a theoretical tunneling spectrum requires the estimation of the interaction of the SE with the electrons, which is quantified in terms of a self-energy, $\Sigma$, derived within MBPT.\cite{SchweflinghausPRB14} 
Once the self-energy is known, the electronic structure can be renormalized and the signature of the SE in the LDOS can be observed.\cite{SchweflinghausPRB14} 
The key equation to be solved is the Dyson equation, that in a schematic notation reads
\bea
  G^\sigma(E) = G_0^\sigma(E) + G_0^\sigma(E) \Sigma^\sigma(E) G^\sigma(E) \;,
  \label{eq:DysonEq}
\eea
where $G_0^\sigma$ corresponds to the initial projected KKR Green function for spin $\sigma\in\{\upa,\dna\}$. 
In its projected form and omitting orbital labels the self-energy is given by
\bea
      \Sigma^{\sigma}(E) &= -\frac{U^2}{\pi}
      \bigg[
               \int_0^\infty \mrm{d}\omega\;
               \Im\!\left[ G_0^{\overline\sigma}(\omega+E)         \overline\chi^{\sigma\overline\sigma}(\omega) \right] \nonumber\\
  &\hspace{1em} - \int_0^{\EF-E} \mrm{d}\omega\;
               \Im\!\left[ G_0^{\overline\sigma}(\omega+E) \right] \overline\chi^{\sigma\overline\sigma}(\omega)^*
      \bigg] \;,
  \label{eq:Sigma}
\eea
where $\overline\chi^{\sigma\overline\sigma}$ is the spherical part of the susceptibility.\cite{SchweflinghausPRB14}

The key statement of the Tersoff-Hamann approximation is that there exists a proportionality between the conductance ($\frac{\partial I}{\partial V}(V)$) as measured in ISTS experiments and the product of LDOS from the tip, $n_\mrm{tip}$, as well as from the probed adatom, $n_\mrm{vac}$, measured at distance $\vc{R}$ away from the adatom,
\bea
          \frac{\mrm{d}I}{\mrm{d}V}(V)
  \propto \left[   n_\mrm{tip}^\upa n_\mrm{vac}^\upa(\EF+V,\vc{R})
                 + n_\mrm{tip}^\dna n_\mrm{vac}^\dna(\EF+V,\vc{R}) \right] \;,\nonumber\\
  \label{eq:TersoffHamann}
\eea
where the densities are directly related to the Green function via
\bea
  n^\sigma(E) = -\frac{1}{\pi} \Im\left[ G^\sigma(E) \right] \;, \quad \sigma\in\{\upa,\dna\} \;.
\eea
The form of Eq.~(\ref{eq:TersoffHamann}) is given such that it accounts for the two spin channels, $\upa$ (majority-spin channel) and $\dna$ (minority-spin channel). 
Thus the renormalization of the electronic structure via Eq.~(\ref{eq:DysonEq}) affords the key ingredient of Eq.~(\ref{eq:TersoffHamann}), the LDOS in vacuum with the contribution of the SEs. 
In this work we assume an $s$-like tip.

With the formalism at hand we concentrate in the following on two analyses. 
In the first part we consider single 3$d$ adatoms on the Pt(111) surface and in the second part we explore the impact of hydrogenation on the SE spectra of Fe and Co adatoms. 

Our method is not established yet to include spin-orbit coupling at the level of the renormalization of the electronic structure.\cite{dosSantosDiasPRB15} 
Thus, we apply an auxiliary field of $B_0=\sunit{10}{\mrm{T}}$ (corresponding to $\sunit{0.58}{\meV}$) that mimics the spin-orbit interaction effect by opening a gap in the SE spectra and allows their comparison on equal footing for the different considered adatoms. 
The resonance of the SE is thus obtained at the Larmor frequency
\bea
  \omega_\mrm{res} = g_\mrm{eff} B_0 \;,
\eea
where $g_\mrm{eff}$ is the effective $g$ value, which usually deviates from 2 in a solid.

\section{Spin excitation of pure 3d adatoms}
\label{sec:PureAdatomSpectra}

\begin{figure}[!t]
 \centering
  \includegraphics[width=0.45\textwidth]{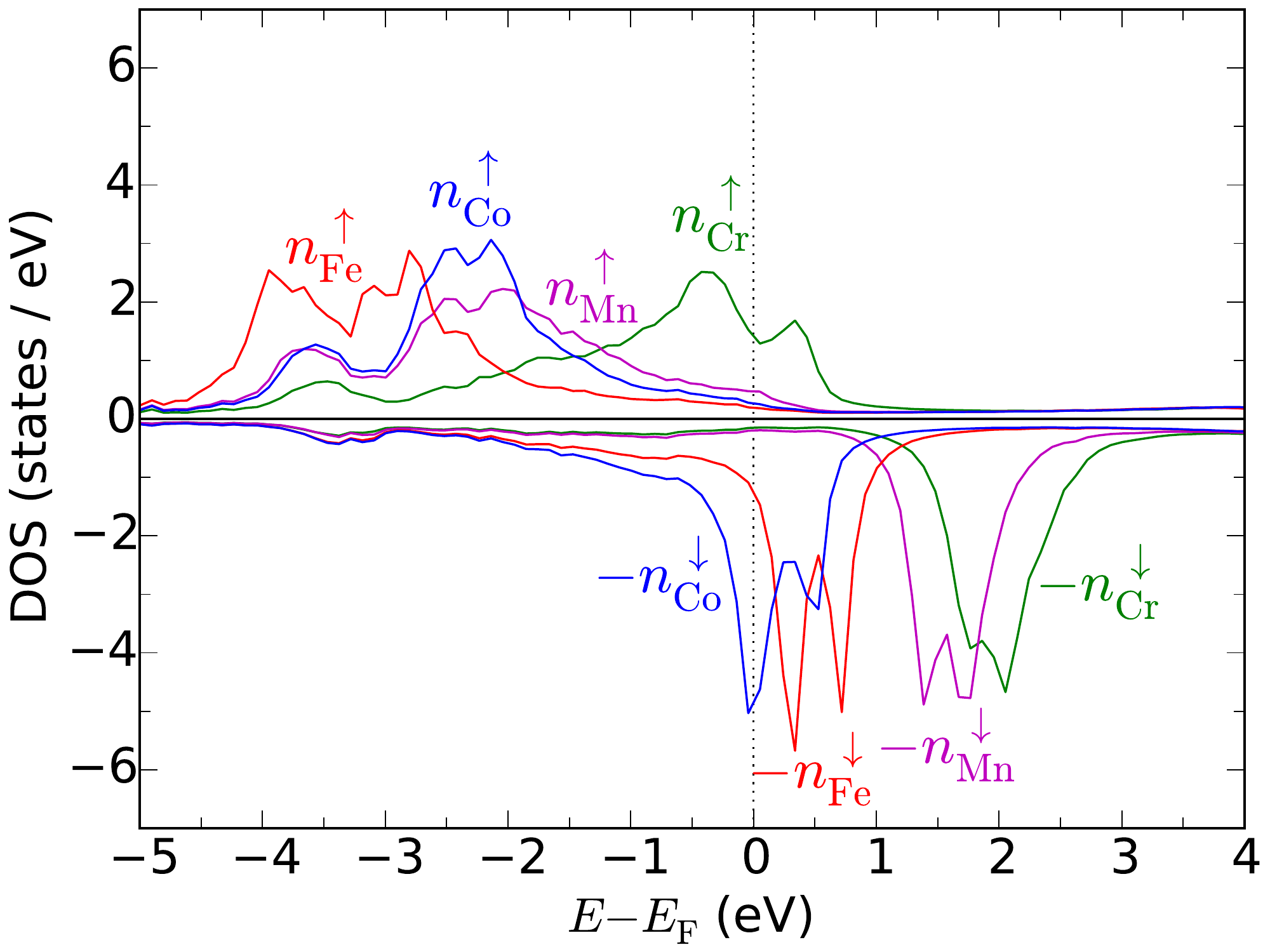}
 \caption{(color online). 
          The spin-resolved total density of states (DOS) are shown 
          for the four adatoms (Cr, Mn, Fe, and Co) placed on the Pt(111) surface. 
          The Fermi energy is indicated by a dotted vertical line.}
 \label{fig:TMPt111_dos}
\end{figure}

In Fig.~\ref{fig:TMPt111_dos} the spin-resolved density of states (DOS) projected onto the adatom is shown. 
In Table~\ref{tab:TMPt111_ChiProperties} we list ground-state properties such as the number of $d$-electrons 
$N_d=\int^{\EF}\text{d}E \left( n_d^\upa(E) + n_d^\dna(E) \right)$ and the spin moment considering only the $d$-orbitals
$m_d=\int^{\EF}\text{d}E \left( n_d^\upa(E) - n_d^\dna(E) \right)$. 
A comparison to the DOS analysis for the adatoms placed on the Cu(111) surface\cite{SchweflinghausPRB14} reveals that the resonances show a stronger splitting on the Pt(111) surface.  
This is due to the hybridization with the $d$-orbitals of the Pt substrate: 
the five $d$-orbitals of the adatoms split into 3 symmetry groups (according to the point group symmetry $C_{3V}$ one has $E_2$ and $E_1$ as linear combinations of $d_{x^2-y^2}$ and $d_{xy}$, as well as $d_{xz}$ and $d_{yz}$, and $A_1$ for the remaining $d_{3z^2-r^2}$). Here, the states located at the Fermi energy 
$\EF$ play a crucial role in determining the observation of possible SEs.\cite{LounisPRB15} 

\begin{figure}[!t]
 \centering
   \includegraphics[trim = 0mm 0mm 0mm 0mm, clip, width=0.45\textwidth]{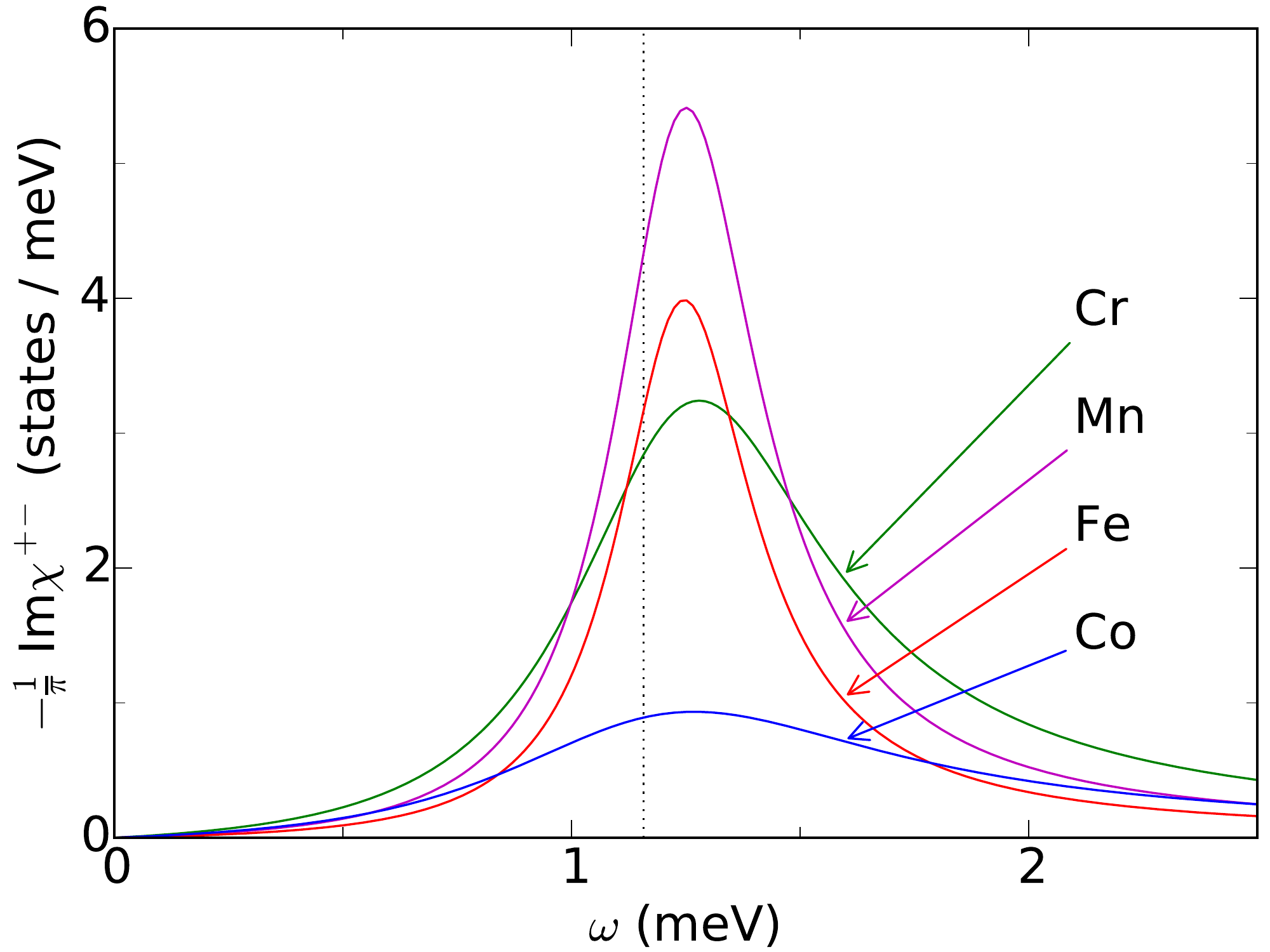}
 \caption{(color online). 
          The imaginary part of the dynamical enhanced susceptibility 
          for the four TM adatoms are shown, describing the intrinsic excitation spectrum. 
          The applied B-field is set to 10 Tesla in order to allow for 
          a comparison among the four different systems. 
          The dotted vertical lines indicate the expected positions of the spin excitations (assuming a $g$ value of $g=2$).}
 \label{fig:TMPt111_Susc}
\end{figure}

The intrinsic magnetic excitation spectrum is given by $\Im\overline\chi^{+-}$, the imaginary part of the enhanced susceptibility. 
In Fig.~\ref{fig:TMPt111_Susc} we show this quantity for the investigated TM adatoms placed at the fcc stacking site on the Pt substrate. 
Depending on the system the spectrum can show a different resonance position which is directly related to the $g$ shift, \ie, the deviation from a $g$ value of exactly 2, which is related to the details of the electronic structure.\cite{LounisPRB15} 
The resonance width is associated with the damping and represents the inverse of the excitation lifetime. 
The origin of the damping is the decay of the SE into Stoner pairs, \ie, electron-hole pairs that propagate in the substrate. 
The decay is mainly driven by available excitation channels in the LDOS at the Fermi energy.\cite{LounisPRB15}
Extracted from these curves, the resonance frequency as well as the $g$ shift and the lifetime are shown in Table~\ref{tab:TMPt111_ChiProperties}. 
The $g$ shift is largest for Co ($g_\mrm{eff}=2.19$) which is of the same order than what was found and measured for the Fe adatom.\cite{KhajetooriansPRL13} 
The largest damping is found for Cr and Co adatoms causing their excitation lifetimes to be three times smaller than those for Mn and Fe adatoms. 
This is rather counter-intuitive for Cr since it is expected to be a half-filled $d$-shell type of adatom. 
However, because of its particular electronic structure induced after deposition on the Pt(111) surface, a rather large DOS in the majority-spin channel is developed leading thereby to a large reservoir for electron-hole excitations.

\begin{table}[!b]
  \begin{ruledtabular}
   \begin{tabular}{lccccc}
    adatom & $N_d$ & $m_d$ ($\mB$) & $\omega_\mrm{res}$ ($\meV$) & $g_\mrm{eff}$ & $\tau_\chi$ ($\mrm{fs}$)     \\
    \midrule
    Cr     & $4.26$ & $3.19$ & $1.28$ & $2.21$ & $494$ \\
    Mn     & $4.97$ & $3.75$ & $1.25$ & $2.16$ & $864$ \\
    Fe     & $5.53$ & $2.79$ & $1.25$ & $2.16$ & $968$ \\
    Co     & $6.91$ & $1.77$ & $1.27$ & $2.19$ & $310$ \\
   \end{tabular}
  \end{ruledtabular}
  \caption{This table shows ground-state properties as well as 
           dynamical properties of the susceptibility for the four adatoms, 
           considering the fcc stacking site of the adatoms. 
           $N_d$ is the number of $d$-electrons and $m_d$ is 
           the magnetic moment when only the $d$-orbitals are considered. 
           For an applied field of $B_0=\sunit{10}{T}$ the resulting 
           effective $g$ value $g_\mrm{eff}=\omega_\mrm{res}/B_0$,
           the resonance position $\omega_\mrm{res}$, 
           and the lifetime $\tau_\chi=\hbar/(2\Gamma)$ of the excitation is shown, 
           where $\Gamma$ is the full-width at half-maximum for the resonances shown in Fig.~\ref{fig:TMPt111_Susc}.}
  \label{tab:TMPt111_ChiProperties}
\end{table}

\begin{figure*}[!t]
 \centering
   \includegraphics[width=0.90\textwidth]{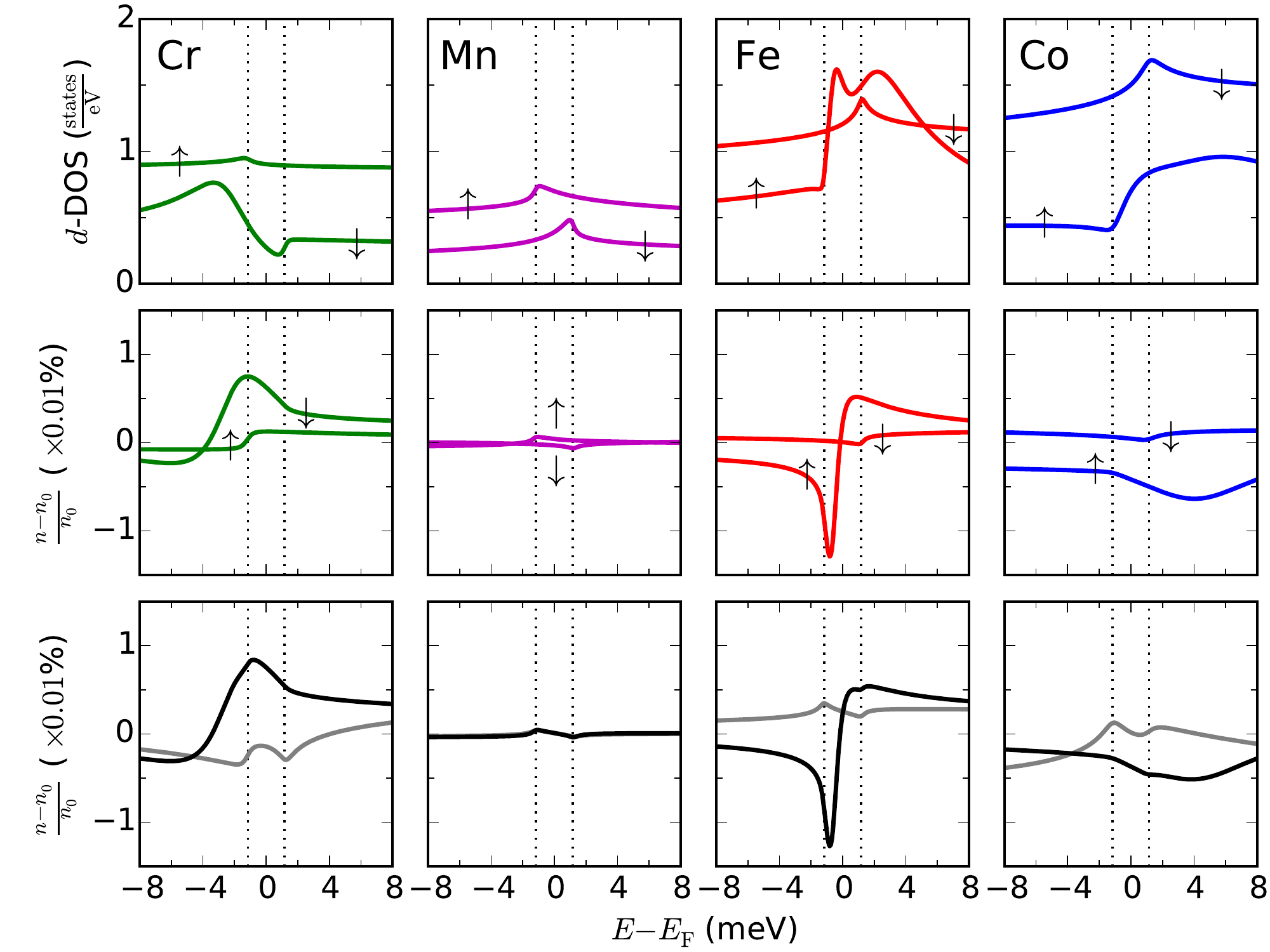}
 \caption{(color online). 
          Renormalized LDOS for TM adatoms (from left to right: Cr, Mn, Fe, Co) on Pt(111). 
          Top row: Renormalized $d$-LDOS in the adatom for majority-spin ($\upa$) and minority-spin ($\dna$) channel. 
          Second row: Efficiency of the theoretical inelastic signal in the vacuum above the adatom calculated from the $s$-LDOS. 
          The efficiency, $(n-n_0)/n_0$, corresponds to the ratio between the change in the DOS ($n-n_0$) due to the SE and the original DOS $n_0$. 
          Third row: Same as second row but summed over spin (black curves), compared to the signal obtained if the electron-SE interaction is not fully considered, \ie, when only the first Born iteration is taken into account (gray curves), for the solution of the Dyson equation, see Eq.~(\ref{eq:firstBorn}). 
          In this case the bound states disappear. 
          The dotted vertical lines are a guide to the eye and indicate the energy 
          at which the excitation resonance is expected, see caption of Fig.~\ref{fig:TMPt111_Susc}.}
 \label{fig:TMPt111_renDOS}
\end{figure*}

After evaluation of the intrinsic SEs, our goal is to explore their impact in the tunneling transport experiment. 
For that we switch from TDDFT to MBPT and solve Eq.~(\ref{eq:DysonEq}) in order to renormalize the LDOS because of the electron-SE interaction. 
The connection to the experimentally measured ISTS spectra is realized via the Tersoff-Hamann approximation\cite{TersoffHamannPRL83} by computing the LDOS in the vacuum region. 
A collection of renormalized LDOS in the energy window around the SE for the four investigated adatoms on Pt(111) are shown in Fig.~\ref{fig:TMPt111_renDOS}. 
The first row of panels represent the impact of the SEs on the LDOS of the adatom, while the second row of panels is devoted to the efficiency of the inelastic excitations in vacuum at the position of the probing tip. 
In addition, the third row shows the spectra as sum over both spin channels as black curves, representing the inelastic spectra as they would be measured in ISTS experiments with a non-polarized tip. 
The efficiency is given by $(n-n_0)/n_0$, where $n_0$ and $n$ represent the LDOS in vacuum before and after the SE is included via the Dyson equation~(\ref{eq:DysonEq}), respectively. 
This permits a comparison of the efficiency of the SEs among the four adatoms and draw conclusions about the possibility of probing them with ISTS. 
The shapes of the calculated spin excitation signatures reveal a great variety of shapes. 
We note that not only steps are observed as a signature of the SEs but also peaks, reversed steps, dips. 
As it was already reported for the same elements placed on the Cu(111) surface,\cite{SchweflinghausPRB14} bound states or satellites are observed in the LDOS, which originate from the interaction of the SE and the electronic structure. 
Indeed, the self-energy of the electrons (not shown), see Eq.~(\ref{eq:Sigma}), provides an additional potential that can create such electronic features. 
For Cr adatom, the minority-spin channel exhibits a satellite around $\sunit{-2}{\meV}$, while for the Fe and Co adatoms such a satellite occurs in the majority-spin channel. 
In contrast to this, the excitation signature in the renormalized LDOS for the Mn adatom exhibits no bound states. 
In the vacuum, the spectra change completely, \eg, in terms of intensity (majority-spin versus minority-spin channel) and shape. 
This observation was already made for adatoms placed on the Cu(111) surface\cite{SchweflinghausPRB14} and is due to the way the electronic states decay into vacuum. 
Thus, the peak in the majority-spin channel of the LDOS for the Cr adatom decays into a reversed step in vacuum, while the bound state present as a resonance in the minority-spin channel below the Fermi energy decays into a reversed step. 
Similar changes occur for all the considered elements. 
Here, it is interesting to note that the efficiency of the Fe inelastic spectra is the highest among all elements. 
The strength of the signature decays further when following the sequence to Cr, Co, and finally Mn. 
This is then in line with the recent experimental observations of Khajetoorians~\etal,\cite{KhajetooriansPRL13,privateComm} who report that the SEs were observed in Fe adatoms but not in Co adatoms. 
Similar findings were reported by Dubout~\etal\cite{DuboutPRL15} 
On top of that one should also keep in mind that we locked the excitation energy at an reference value, while it is known that the excitation energy depends on the system: 
for Co adatoms the anisotropy is about 10 times larger than the one for Fe adatoms\cite{BalashovPRL09,KhajetooriansPRL13,DuboutPRL15} and thus, an even larger broadening of the excitation signature reduces further the probability of observing the SEs in Co adatoms. 
As far as the other elements are concerned our analysis allows to the following prediction: 
on the one hand Cr adatoms could be a good candidate to observe the signature of a SE in ISTS measurements, as the signature in Fig.~\ref{fig:TMPt111_renDOS} is in the order of the one for Fe adatoms. 
On the other hand the efficiency for the SE for Mn adatoms is the least pronounced among the investigated adatoms and therefore the observation of the SE within an ISTS experiment is unlikely. 

At this stage, it is important to point out that the lifetime and energy of the SE extracted from the intrinsic excitation spectra discussed in the previous section can be different from those obtained from the renormalized electronic structure. 
This is exemplified in the Appendix for the case of the Fe adatom. 

Up to now, we considered the full electron-SE interaction by solving the Dyson equation, see Eq.~(\ref{eq:DysonEq}). 
It is interesting however to analyze the outcome of the first order Born iteration in solving that equation.
\bea
  G_\mrm{1stB}^\sigma(E) = G_0^\sigma(E) + G_0^\sigma(E) \Sigma^\sigma(E) G_0^\sigma(E) \;.
  \label{eq:firstBorn}
\eea
The difference is the replacement of $G$ by $G_0$ at the right-hand side of Eq.~(\ref{eq:DysonEq}). 
If the self-energy is a small perturbation we expect the result from the first Born iteration to be a good approximation to the full solution. 
If the self-energy is strong, for example, if an extra resonance develops, the first Born approximation is no longer adequate. 
A comparison of the renormalized electronic structure in the vacuum is given in the third row of Fig.~\ref{fig:TMPt111_renDOS}. 
The gray curves indicate the resulting LDOS, summed over spin, when Eq.~(\ref{eq:firstBorn}) is utilized, whereas the black curves indicate the same quantity when the full Dyson equation (see Eq.~(\ref{eq:DysonEq})) is solved. 
Again, we observe different behaviors for the four cases: 
whereas for the Mn adatom the first order Born iteration is already sufficient to capture the full renormalization of the electronic structure, other systems (such as Cr or Fe adatoms) only establish bound states (satellite) for higher orders in the scattering series (\ie, beyond the first order Born iteration). 
This is an important conclusion as it shows in certain cases the utilization of the full Dyson equation is unavoidable when one is interested to find the proper renormalization of the electronic structure.

In the next Section we focus on the object of the experimental controversy: Fe and Co adatoms by exploring their hydrogenation by investigating first the placement of hydrogen.

\section{Spin excitation of hydrogenated F\lowercase{e} and C\lowercase{o} adatoms}
\label{sec:HydrogenatedAdatomSpectra}

\subsection{Structure optimization: Where does hydrogen go?}
\label{sec:TMPt111_FeHdimer_StructureOptimization}

\begin{figure*}[!t]
 \centering
   \includegraphics[trim = 5mm 5mm 0mm 35mm, clip, width=0.90\textwidth]{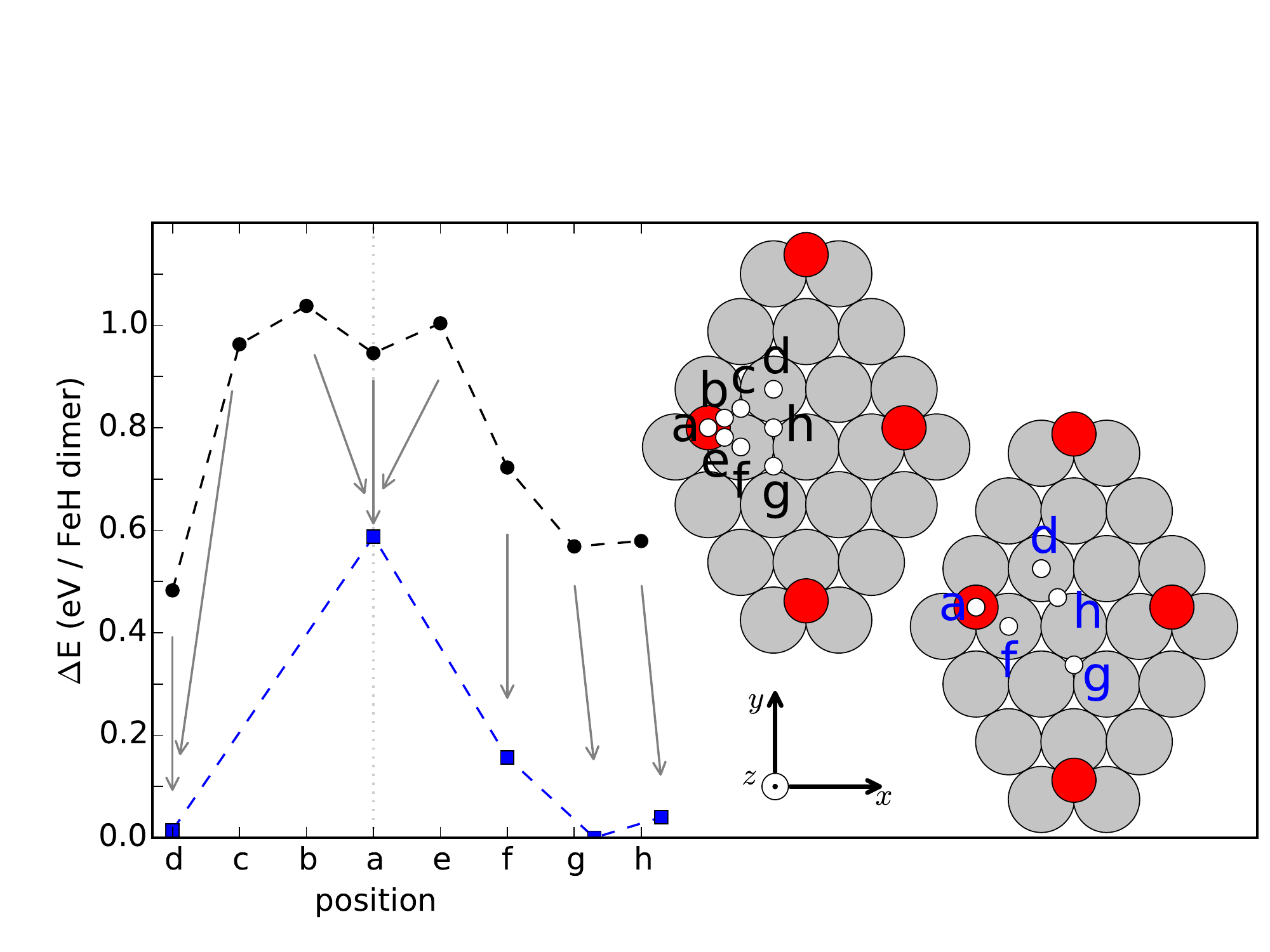}
 \caption{(color online). 
          The total energies for the 3$\times$3 supercell are shown 
          with respect to eight different starting positions for H. 
          The top and bottom insets on the right-hand side represent possible $\mrm{FeH}$ dimers, 
          labeled from \textit{a} to \textit{h}. 
          Different constraints in the performed relaxations are considered: 
          Black circles indicate relaxation of the $\mrm{FeH}$ dimer along the $z$-direction, 
          while blue squares corresponds to a full relaxation of the dimer and the surface Pt layer.}
 \label{fig:energy_vs_position}
\end{figure*}

In order to perform reasonable simulations of the inelastic spectra measured with ISTS, we first investigate different configurations of $\mrm{FeH}$ dimers placed on the Pt(111) surface and compare the relaxed setups in terms of the resulting total energies by use of the VASP program. 
In Fig.~\ref{fig:energy_vs_position} the total energies are plotted with respect to the initial position of the H atom, for the fcc stacking of the Fe atom. 
The total energies are given with respect to the global minimum value (relaxed position when starting with position \textit{g}, blue squares), which enables a direct comparison of their relative values. 
Different constraints were considered in the performed relaxations: 
The calculations represented by the black circles indicate that the impurity atoms are relaxed along the $z$-direction only while the Pt atoms are kept fixed in bulk-like positions. 
The blue squares were obtained after allowing more degrees of freedom, where the impurity atoms and Pt atoms of the surface layer nearby are relaxed in all three directions. 
In addition to the curves the figure contains two small top view sketches of the corresponding supercell, where the gray circles and the red circle correspond to Pt atoms and the Fe adatom, respectively. 
The small white circles indicate different positions of the H atom with their respective label, when the $\mrm{FeH}$ dimer is relaxed along the $z$-direction only (top inset, black labels) and when the dimer and the surface Pt layer are fully relaxed (bottom inset, blue labels). 
Note, that the label colors black and blue match the color code used for the energy curves.

The black circles in Fig.~\ref{fig:energy_vs_position}  are connected with dashed lines that act as a guide to the eye and reveal two important outcomes: 
on the one hand, the H atom shows the tendency to nestle at the side of Fe rather than on top of it. 
The total minimum of the investigated structures appears for position \textit{d}, the structural arrangement where H is on top of a Pt atom and as far as possible separated from the neighboring Fe impurities. 
On the other hand, placing H on top of Fe (position \textit{a}) reveals a local minimum (position \textit{b} and \textit{e} are nearby but higher in energy) and that acts then as a trap for any additional H atom.

By relaxing the impurity atoms as well as the Pt surface layer in all three directions (blue squares) we notice that for some arrangements the H atom travels some distance within the supercell (usually not more than $\sunit{2}{\Angs}$) whereas the Fe atom's position remains unchanged (below 1\%). 
The relaxation calculations for position \textit{a}, \textit{b}, and \textit{e} lead to the same structural arrangement, a local minimum on top of Fe (position \textit{a}), indicated by the equal values for their total energies (see blue curve in Fig.~\ref{fig:energy_vs_position}). 
For position \textit{c} we obtain a large energy gain (above $\sunit{0.4}{\eV}$ per $\mrm{FeH}$ dimer) compared to the value for the black curve. 
In fact the structure now matches the one from position \textit{d}, an indication that the H atom prefers to be located on top of a Pt atom rather than close to the adatom. 
This gets confirmed by the fact that the H atom at position \textit{f} does not move away from its position on top of Pt even though it starts with the same distance to the Fe atom as it is the case for position \textit{c}. 
Finally, we state that the largest energy gain for all investigated structures (with respect to the black data points) is achieved through the relaxations of the Pt atoms of the topmost layer, as could be verified by exclusively relaxing these atoms alone. 
In conclusion the performed structural optimization calculations reveal the following three main conclusions: 
($i$)   Hydrogen shows the preference to be located further apart from the Fe adatom. 
($ii$)  In the examined energy landscape of positioning the $\mrm{FeH}$ dimer, one observes a local minimum for having the H atom located on top of the Fe atom. 
($iii$) A global minimum is found when the H atom is in contact with the Pt surface layer.  
These statements illustrate the structural optimization analysis for the hydrogenation of the Fe adatom. 
Test calculations for the same setups with Co as adatom have been performed as well and allow to draw the same conclusions on a qualitative level as summarized above for the Fe adatom. 
From these simulations we believe that in the experimental conditions of hydrogenation, there is a low but finite probability of having a H atom attached to adatom.

\subsection{Effect of hydrogen on spin excitation spectra}
\label{sec:FeH_CoH_spectra}

\begin{figure}[!t]
 \centering
  \begin{tikzpicture}[scale=1.0]
   \node (0,0) {\includegraphics[width=0.45\textwidth]{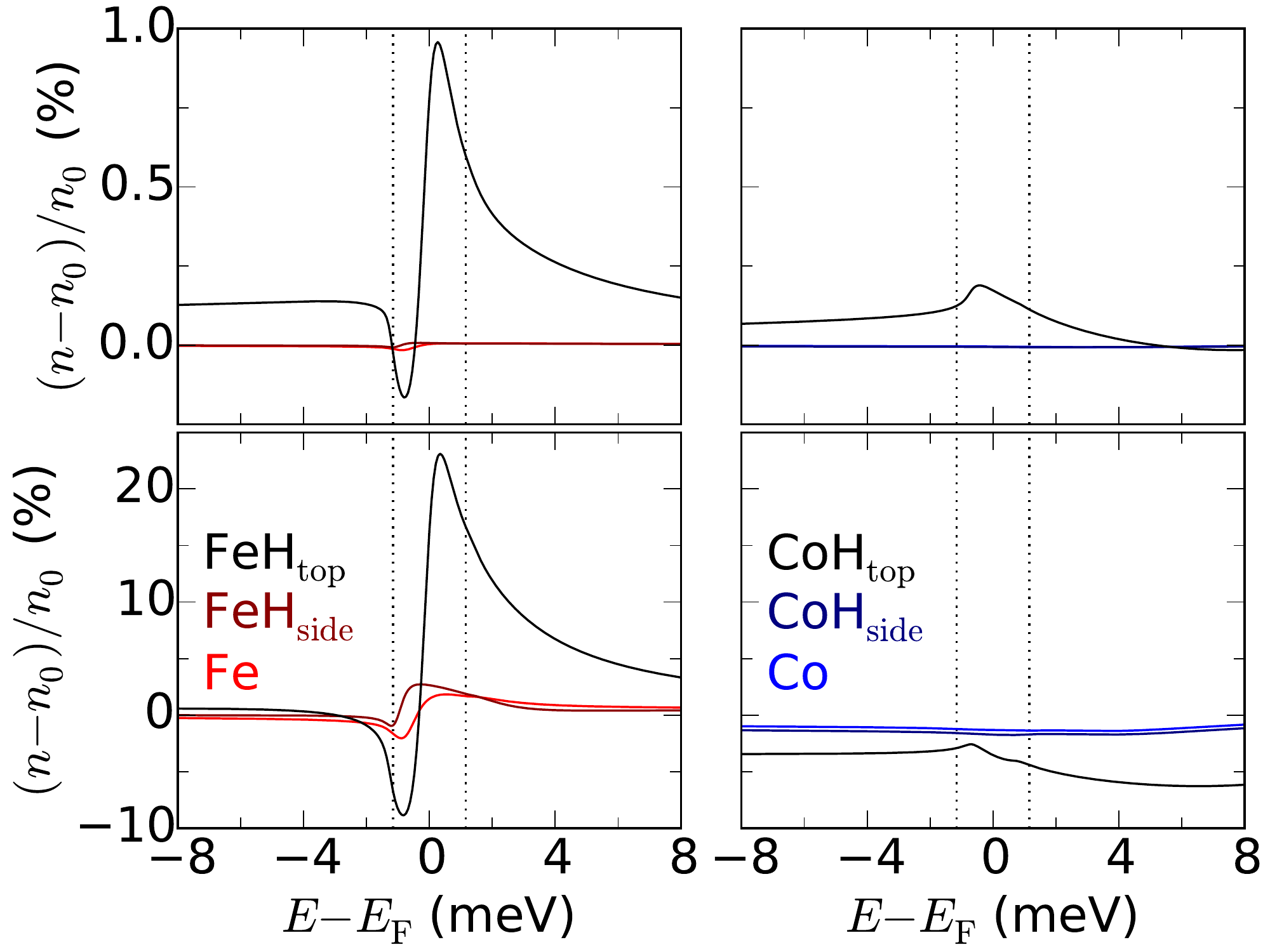}};
   \node at (-2.5,+2.5) {(a)};
   \node at (+1.1,+2.5) {(b)};
   \node at (-2.5,-0.1) {(c)};
   \node at (+1.1,-0.1) {(d)};
  \end{tikzpicture}
 \caption{(color online). 
          This figure illustrates the impact of hydrogen on the 
          magnetic excitation spectra, investigating the efficiency of the inelastic signal at 
          (a) $\sunit{6.8}{\Angs}$ away from the Fe adatom 
          and (b) the Co adatom, and at 
          $\sunit{4.8}{\Angs}$ away from the adatom 
          ((c) and (d) for Fe and Co adatom, respectively). 
          For each case the shape of the spectrum for the pure adatom 
          is compared to those for adatoms that are contaminated by 
          one H atom at the side ($H_\mrm{side}$) or an top of it ($H_\mrm{top}$). 
          The dotted vertical lines indicate the expected positions of the 
          SEs (assuming $g=2$).}
 \label{fig:TMPt111_FeH_CoH_renDOS}
\end{figure}
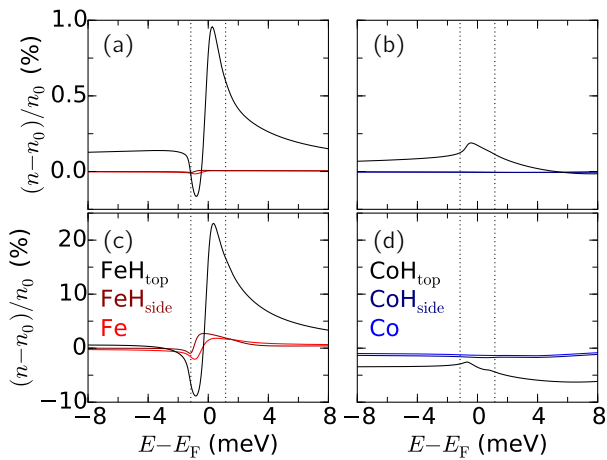

The structural analysis of Sec.~\ref{sec:TMPt111_FeHdimer_StructureOptimization} reveals two possible scenarios of where hydrogen could be located: 
either it is adsorbed in the layer above the adatom, where a local energy minimum was found, or it directly sits on the Pt surface next to the adatom. 
Although the latter is lower in energy the following analysis accounts for both scenarios, since both are considered as being realized $\mrm{FeH}$ or $\mrm{CoH}$ dimers in experimental measurements. 
In the present analysis the simulations of the SEs are carried out by means of the KKR formalism, which allows for two different arrangements of $\mrm{FeH}$ and $\mrm{CoH}$ dimers that are most similar to those suggested by the structural relaxation analysis from the previous Sec.~\ref{sec:TMPt111_FeHdimer_StructureOptimization}: 
($i$)  a dimer where H is located at $\vc{R} = \left(\frac{1}{\sqrt{2}},0,0\right) a$ away from the adatom, where $a$ is the lattice constant as defined in Sec.~\ref{sec:Method} (in the following referred to as $\mrm{FeH}_\mrm{side}$ and $\mrm{CoH}_\mrm{side}$) and 
($ii$) a dimer where H is located at $\vc{R} = \left(0,\frac{1}{\sqrt{6}},\frac{1}{\sqrt{3}}\right) a$, in the layer above the adatom (in the following referred to as $\mrm{FeH}_\mrm{top}$ and $\mrm{CoH}_\mrm{top}$). 
These positions of H are most comparable with positions \textit{h} and \textit{f} as given in the top inset of Fig.~\ref{fig:energy_vs_position}. 
The vacuum sites at which the SEs are probed are located in the second layer above the adatom (with $\vc{R} = \left(0,-\frac{1}{\sqrt{6}},\frac{2}{\sqrt{3}}\right) a$, referred to as $\mrm{vc2}$) and in the third layer above the adatom (with $\vc{R} = \left(0,0,\sqrt{3}\right) a$, referred to as $\mrm{vc3}$). 
Thus, the sites at $\mrm{vc2}$ and $\mrm{vc3}$ correspond to a distance of $\sunit{4.8}{\Angs}$ and $\sunit{6.8}{\Angs}$ to the adatom, respectively. 

Since experimentally a focus is set on Fe and Co adatoms (\eg, Refs.~\onlinecite{BalashovPRL09}, \onlinecite{KhajetooriansPRL13}, and \onlinecite{DuboutPRL15}) we concentrate on how a side position or a top position of hydrogen changes their excitation spectra. 
Following the same procedure as presented in the previous Sec.~\ref{sec:PureAdatomSpectra} we arrive at the renormalized LDOS spectra in two different vacuum positions above the adatom. 
Like before, a universal auxiliary magnetic field of $B_0=\sunit{10}{\mrm{T}}$ mimics a gap induced by spin-orbit coupling (SOC) in the excitation spectra.

In Fig.~\ref{fig:TMPt111_FeH_CoH_renDOS} the impact of hydrogenation on the efficiency of the SEs on the tunneling spectra of Fe adatoms and Co adatoms is shown, probed at two different vacuum sites. 
For the Fe systems (see Figs.~\ref{fig:TMPt111_FeH_CoH_renDOS}(a) and \ref{fig:TMPt111_FeH_CoH_renDOS}(c)) the presence of H at the side of Fe sharpens the resonance widths meaning that the lifetime of the excitation is enlarged. 
For the top position of the H atom, the satellite (coming from the spin-up channel) moves to the Fermi energy and interferes with the excitation signature originating from the spin down channel (expected at about $\sunit{+1.1}{\meV}$). 
For the Co adatom (see Figs.~\ref{fig:TMPt111_FeH_CoH_renDOS}(b) and \ref{fig:TMPt111_FeH_CoH_renDOS}(d)) the side position of the H atom does not affect the excitation spectra in a remarkable way. 
When it is placed on top of the Co adatom, however, an interesting change in the renormalized spectra is observed: 
Near the Fermi energy a resonance appears and dominates the shape of the spectra completely compared to the spectrum of the Co adatom alone.

We also notice that for both impurities the top position of hydrogen leads to an enhanced efficiency compared to the pure adatom system or the one where hydrogen is set at the side next to the adatom, within the same layer. 
A possible reason for this can be traced back to the symmetry of the investigated structures. 
The coupling between the $s$-like tip in vacuum and the adatom $d$-orbitals is contained in the connecting Green function $G_0^\mrm{adatom,vc}$. 
For the pure adatom one has the point group symmetry $C_{3V}$, for which only the $d_{z^2}$ off-diagonal element leads to a coupling to the $s$-like tip by symmetry. 
The presence of hydrogen in the system lowers the symmetry, such that other orbital elements lead to a coupling to different orbital components of the self-energy. 
Thus additional inelastic tunneling channels open up, which can enhance the magnitude of the theoretical inelastic spectra. 
For the $\mrm{H}_\mrm{side}$ setups the connection between the $s$-like tip and the $d$-orbitals other than $d_{z^2}$ are usually two orders of magnitude smaller than the one connecting the $d_{z^2}$ orbital of the adatom to the $s$ orbital in the vacuum. 
Thus, the resulting renormalized spectra are not much affected. 
This is different for the $\mrm{H}_\mrm{top}$ setups, where the $s-d_{xy}$ and $s-d_{yz}$ elements of the connecting Green function 
are of the same order of magnitude than the $s-d_{z^2}$ connecting element. 
This is a strong indication that the presence of H can affect measured inelastic excitation spectra substantially, a conclusion that is in line with recent ISTS experiments.\cite{DuboutPRL15} 
In Figs.~\ref{fig:TMPt111_FeH_CoH_renDOS}(c) and \ref{fig:TMPt111_FeH_CoH_renDOS}(d) the same setups are investigated for the vacuum site $\mrm{vc2}$ that is one layer closer to the surface structure. 
The figures illustrate that the shapes of the two spectra is not affected much by bringing the tip closer to the adatom. 
However, one observes that the efficiency increased compared to the efficiency shown in Figs.~\ref{fig:TMPt111_FeH_CoH_renDOS}(a) and \ref{fig:TMPt111_FeH_CoH_renDOS}(b), where the vacuum site $\mrm{vc3}$ is probed. 
For the case of $\mrm{FeH}_\mrm{top}$ one even observes nearly 25\% efficiency and thus a high probability of measuring the SEs. 
The signatures when placing hydrogen on top of Fe or on top of Co do not scale with a factor of 100 anymore compared to the other two setups (pure Fe and Co as well as Fe and Co with the hydrogen at the side). 
Instead the difference is reduced to a factor of 10. 
If one assumes that in experiment the distance of the tip to the probing nanostructure is varied until a signal is observed, one can conclude that depending on the distance to the probed nanostructure a SE signal is observable or not. 
Thus, the efficiency depends not only on the probed adatom and the hydrogenation, but also on the distance of the STM tip to the nanostructure.

\begin{figure}[!t]
 \centering
  \begin{tikzpicture}[scale=1.0]
   \node (0,0) {\includegraphics[width=0.45\textwidth]{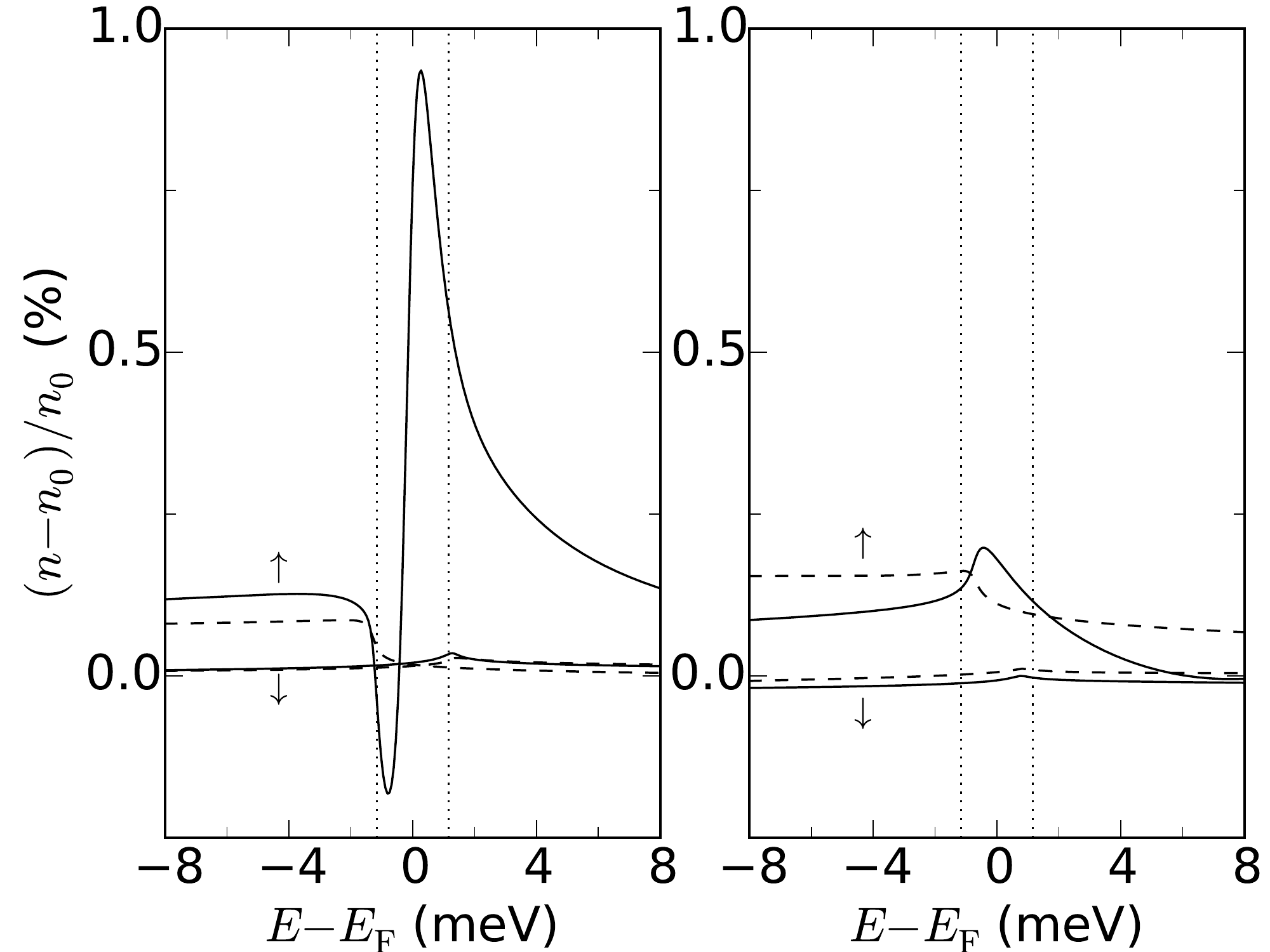}};
   \node at (-2.7,+2.5) {(a)};
   \node at (+1.0,+2.5) {(b)};
  \end{tikzpicture}
 \caption{(color online). 
          For the two systems (a) $\mrm{FeH}_\mrm{top}$ and (b) $\mrm{CoH}_\mrm{top}$ this figure 
          shows a comparison of the efficiency of the inelastic signal 
          in the vacuum when the first order Born iteration (see Eq.~(\ref{eq:firstBorn}))
          is used to the efficiency obtained when the Dyson equation (see Eq.~(\ref{eq:DysonEq})) is solved. 
          The majority-spin channel of Fe ($\uparrow$) shows a bound state 
          that only appears when the Dyson series is summed to all orders. 
          For the other spin channel ($\downarrow$) and the two spin 
          channels of the Co system the first order Born iteration is able to capture 
          most of the inelastic features.}
 \label{fig:TMPt111_FeH_CoH_renDOS_fb}
\end{figure}

Finally we show in Fig.~\ref{fig:TMPt111_FeH_CoH_renDOS_fb} the impact of solving the first order Born iteration, see Eq.~(\ref{eq:firstBorn}), and compare it to the result when the full Dyson equation is solved, see Eq.~(\ref{eq:DysonEq}). 
The focus here is set on the dimers of $\mrm{FeH}$ and $\mrm{CoH}$ where the hydrogen atom is set in the layer above the adatom measured at the vacuum site $\mrm{vc3}$. 
The largest difference appears for the majority-spin channel of the Fe system, where the clear step in the first order Born iteration is not present in the full solution of the Dyson equation. 
One sees that the resonance representing a bound state (slightly below the Fermi energy) is dominant and can alter the electronic structure around the Fermi energy substantially. 
The minority-spin channel is nearly unaffected. 
The same is true for the two spin channels of the $\mrm{CoH}_\mrm{top}$ system.

\section{Summary}
\label{sec:Summary}

In this paper we studied spin excitations in TM adatoms on the Pt(111) surface utilizing a newly developed method that accounts for the interaction of the spin excitation with the electronic structure.

In a first part, the calculated excitation spectra for Cr, Mn, Fe, and Co adatoms were analyzed. 
The analysis incorporated the presentation and discussion of intrinsic excitation spectra (provided by the susceptibility) as well as the signature of the spin excitation in the renormalized local density of states in the adatom and the vacuum above the adatom. 
In the spirit of the Tersoff-Hamann approximation\cite{TersoffHamannPRL83} the latter is proportional to the experimentally measured inelastic spectra. 
We notice that the shape of the calculated excitation spectra may deviate from expected step-like shapes and bound states (satellite) may occur, as it was already found for 3$d$ adatoms on the Cu(111) surface.\cite{SchweflinghausPRB14} 
We predict the presence of spin-excitations signature in the four types of investigated adatoms but the efficiency of the inelastic signals is found to be strongly dependent on the chemical nature of the deposited atoms and decreases in intensity following the sequence Fe, Cr, Co and Mn. Thus, future ISTS measurements with higher energy resolution and lower temperature than what is available today may be able to measure the predicted inelastic spectra.

In a second part, we focused on the impact of hydrogenation on the obtained excitation spectra of Fe and Co adatoms. 
A structural optimization analysis revealed two possible scenarios: 
either H is positioned on top of the adatom in a local minimum, or it is adsorbed on the Pt surface next to the adatom. 
Accounting for both absorption scenarios of the H atom, the excitation spectra for Fe and Co adatoms were calculated and compared to the shape of the spectra obtained for the same systems without H. 
In accordance to previously reported ISTS measurements\cite{DuboutPRL15} the shape can show a strong dependence on the hydrogenation of the structure. 
Most importantly, these findings reveal that the ability to measure spin excitation may be strongly affected by the presence or absence of hydrogen, its position in the setup, the position of the tip. 
This can be used to explain controversial findings that are reported in the literature.

\begin{acknowledgments}
We acknowledge fruitful discussions with P.~H.~Dederichs, S.~Bl\"ugel, S.~Lichtenstein, M.~Valentyuk, A.~A.~Khajetoorians, and J.~Wiebe. 
Research supported by the HGF-YIG Programme VH-NG-717 (Functional nanoscale structure and probe simulation laboratory -- Funsilab).
\end{acknowledgments}

\appendix

\section*{Appendix}

\begin{figure}[!b]
 \centering
  \begin{tikzpicture}[scale=1.0]
   \node (0,0) {\includegraphics[trim = 0mm 0mm 0mm 0mm, clip, width=0.45\textwidth]{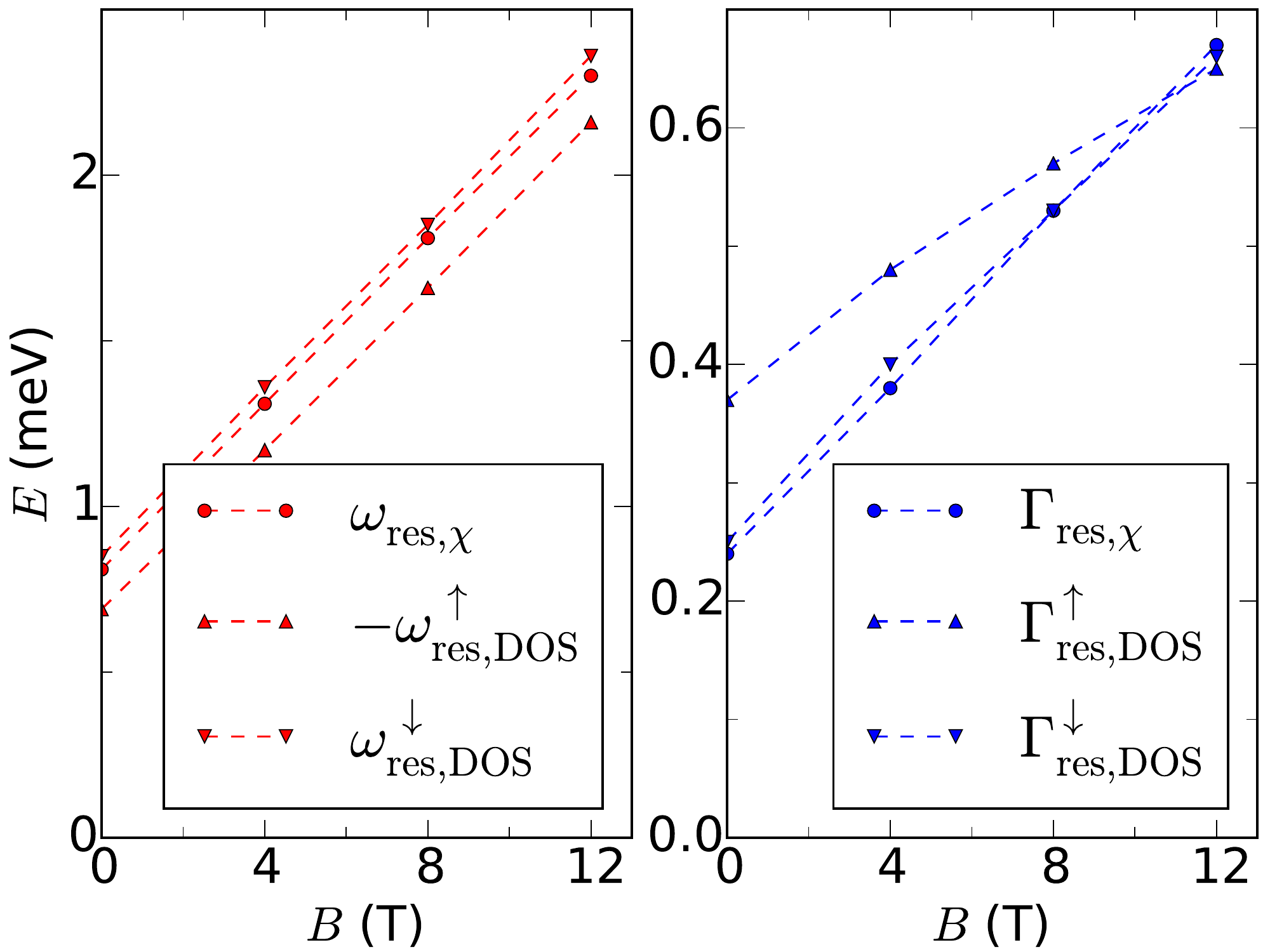}};
   \node at (-3.0,+2.5) {(a)};
   \node at (+1.0,+2.5) {(b)};
  \end{tikzpicture}
 \caption{(color online). 
          Fe adatoms on Pt(111). 
          (a) The resonance frequency, $\omega_{\mathrm{res},\chi}$, as well as 
          (b) the full-width at half-maximum, $\Gamma_{\mathrm{res},\chi}$, 
          using the intrinsic excitation spectra (given by the susceptibility $\chi$) 
          are compared to the spin-resolved values when using the renormalized DOS. 
          Note that $\omega_{\mathrm{res},\mathrm{DOS}}^\upa$ corresponds to 
          the step at negative energy and is plotted with a minus sign 
          which allows a comparison among all three resonances. 
          In order to mimic the SOC-induced gap found in Ref.~\onlinecite{KhajetooriansPRL13} 
          an auxiliary magnetic field of $B_0=\sunit{0.375}{\meV/\mB}$ is assumed, 
          see Fig.~4 related to Fe sitting on fcc stacking site of Ref.~\onlinecite{KhajetooriansPRL13}.}
 \label{fig:TMPt111_Fe_res_FWHM}
\end{figure}

It is instructive to compare the properties of the intrinsic spin-excitations (given by the imaginary part of the susceptibility $\chi$) 
and their signature in the electronic structure as measured within theoretical ISTS (given by the renormalization of the DOS via the self-energy $\Sigma$). 
In  Fig.~\ref{fig:TMPt111_Fe_res_FWHM}, we compare the energy and linewidths (inverse lifetimes) of the spin-excitations for a pure Fe adatom on Pt(111) surface 
as function of an applied magnetic field. An auxiliary magnetic field of $B_0=\sunit{0.375}{\meV/\mB}$ mimics the SOC-induced gap in the ISTS measurements from 
Ref.~\onlinecite{KhajetooriansPRL13}. These properties follow a linear increase with the magnetic field but one notices a discrepancy between the intrinsic spin-excitations 
and ISTS spectra. For instance, the linewidths of the ISTS spectra are larger than those obtained from the intrinsic spectra. The lifetime of the resonance in the intrinsic excitation spectrum drops from about $\sunit{1370}{\mrm{fs}}$ (for zero $B$-field) to $\sunit{491}{\mrm{fs}}$ (for $\sunit{12}{T}$). Moreover, there is a spin-asymmetry in the 
linewidths and the excitation energies, which is related to the spin-asymmetry of the DOS of Fe around the 
Fermi energy.  Also, we notice that the presence of the satellite in the spin up channel (\cf~Fig.~\ref{fig:TMPt111_renDOS}) is located near the Fermi energy and therefore strongly affects the shape of the calculated inelastic spectra. 
Thus, this satellite leads to a stark distortion of resonance and linewidth of the minority-spin channel when the renormalized spectra of the two spin channels are superimposed with equal weights.

\bibliography{references}

\end{document}